\documentclass[conference]{IEEEtran}
\IEEEoverridecommandlockouts
\usepackage{comment}
\usepackage{cite}
\usepackage{amsmath,amssymb,amsfonts}
\usepackage{algorithmic}
\usepackage{graphicx}
\usepackage{textcomp}
\usepackage{xcolor}
\usepackage{booktabs}
 \usepackage{multirow}
 \usepackage{url}
\def\BibTeX{{\rm B\kern-.05em{\sc i\kern-.025em b}\kern-.08em
    T\kern-.1667em\lower.7ex\hbox{E}\kern-.125emX}}

\begin{document}

\title{STAMP-2.5D: Structural and Thermal Aware Methodology for Placement in 2.5D Integration
\thanks{This material is based upon work supported by the PRISM Center under the JUMP2.0 Program sponsored by Semiconductor Research Corporation (SRC)}}

 \author{
     \IEEEauthorblockN{
         Varun Darshana Parekh\IEEEauthorrefmark{1}, Zachary Wyatt Hazenstab\IEEEauthorrefmark{1}, Srivatsa Rangachar Srinivasa\IEEEauthorrefmark{2}, \\
         Krishnendu Chakrabarty\IEEEauthorrefmark{3}, Kai Ni\IEEEauthorrefmark{4}, Vijaykrishnan Narayanan\IEEEauthorrefmark{1}
     }
     \IEEEauthorblockA{
         \IEEEauthorrefmark{1}The Pennsylvania State University, USA
         \IEEEauthorrefmark{2}Intel, Hillsboro, USA\\
         \IEEEauthorrefmark{3}Arizona State University, USA
         \IEEEauthorrefmark{4}University of Notre Dame, USA  
     }
    }

\maketitle

\begin{abstract}
Chiplet-based architectures and advanced packaging have emerged as transformative approaches in semiconductor design. While conventional physical design for 2.5D heterogeneous systems typically prioritizes wirelength reduction through tight chiplet packing, this strategy introduces thermal bottlenecks and intensifies coefficient of thermal expansion (CTE) mismatches, compromising long-term reliability. Addressing these challenges requires holistic consideration of thermal performance, mechanical stress, and interconnect efficiency. We introduce STAMP-2.5D: Structural and Thermal Aware Methodology for Placement in 2.5D integration, the first automated placement methodology that simultaneously optimizes these critical factors. Our approach employs finite element analysis to simulate temperature distributions and stress profiles across chiplet configurations while minimizing interconnect wirelength. Experimental results demonstrate that a thermal and structurally aware automated placement approach reduces overall stress by 11\%, 
maintains excellent thermal performance with a negligible 0.5\% temperature increase and simultaneously reduces total wirelength by 11\% compared to temperature-only optimization. Additionally, we conduct an exploratory study on the effects of temperature gradients on structural integrity, providing crucial insights for reliability-conscious chiplet design. 
\end{abstract}

\vspace{-0.4 cm}
\section{Introduction}
The continuous advancement of semiconductor technology has driven an ever-increasing demand for higher performance, greater functionality, and enhanced integration density in electronic systems. Chiplet-based architectures, particularly 2.5D integration, have emerged as promising solutions in the electronics industry, bridging the gap between chip and package sizes by mounting multiple dies on a single package substrate. This approach enhances capacity and performance by facilitating the heterogeneous integration of chiplets with varied functionalities and material properties into unified systems~\cite{Zhou2022}. Advanced packaging technologies, including Through-Silicon-Via (TSV) fabrication methods and multi-level assembly techniques, have further solidified the reliability and affordability of 2.5D packaging~\cite{Shao2018}.

Recent advancements in chiplet architectures and advanced packaging technologies have revolutionized system design, offering viable pathways to address the slowdown of Moore's Law. 
Commercial implementations leverage chiplet architecture to integrate 7nm CPU dies with 12nm I/O dies for optimal performance and cost~\cite{Naffziger2020,Naffziger2021}. 
Further advances demonstrate scalability and efficiency in chiplet-based systems using heterogeneous 2.5D packaging techniques for workload-dependent configurations~\cite{srivatsa25}. 
These innovations have catalyzed diverse integration strategies, including silicon interposers, bridge-chip technology, active interposers, and 3D interconnects~\cite{Zhang2017,Zhou2022,Coudrain2019}. Chiplet integration thus enables scalable architectures, flexibility in node selection, and 
reuse of off-the-shelf intellectual properties (IP) blocks\cite{Kim2020,Mounce2016}. However, substantial challenges persist, notably in inter-die communication, mixed-node integration, and thermal management\cite{Loh2021,Han2022,10900656}.

Thermal management has become increasingly critical in 2.5D chiplet systems
 due to higher integration and power densities. One such approach, TAP-2.5D, strategically inserts spacing between chiplets to jointly optimize thermal performance and wirelength~\cite{TAP25D}. Extensive research efforts employing finite element analysis and computational fluid dynamics have studied thermal behaviors and factors influencing package warpage, die stress, and solder joint reliability~\cite{Wang2018,Shao2018,Zhang2017}. Nevertheless, while thermal considerations have been extensively explored, mechanical stress implications, particularly arising from the mismatch in the Coefficient of Thermal Expansion (CTE) among heterogeneous dies and interposers, remain inadequately investigated. \textcolor{black}{Beyond CTE-driven thermo-elastic stress, large 2.5-D assemblies are also subject to the self-weight of stacked chiplets, heat-spreaders, HBM modules and copper planes, which induces non-uniform bending and warpage across the interposer. In addition, residual stresses locked in during wafer thinning, die-attach curing and under-fill solidification can modify the absolute stress values as well. Because these mechanical loads differ in both magnitude and spatial distribution from the temperature field,} thermomechanical stress poses substantial reliability risks in 2.5D packages, warranting integrated consideration alongside thermal and interconnect optimization~\cite{Kim2023}. 

\begin{figure}[h!]
    \centering
    \includegraphics[width=0.7\linewidth]{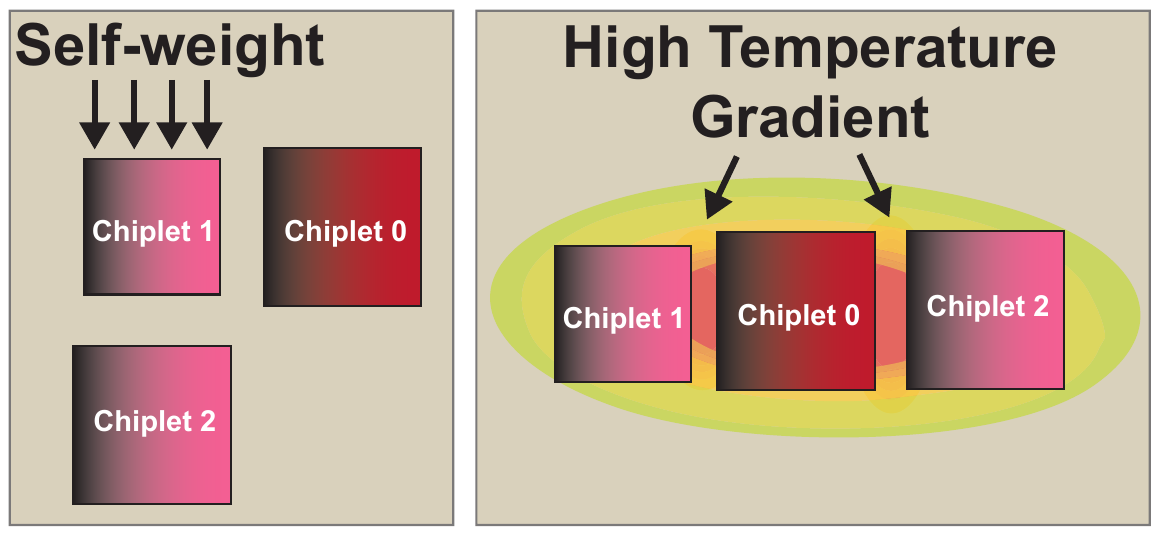}
    \caption{ (left) Uniform-temperature placement with self-weight bending stress; (right) symmetric placement with thermo-elastic stress from temperature gradients.}
    \vspace{-0.2cm}
    \label{fig:toy_example}
    \vspace{-0.5cm}
\end{figure}

\textcolor{black}{Fig.~\ref{fig:toy_example} provides a simplified illustration of these two distinct stress drivers. In left panel, a uniform temperature field combined with asymmetric chiplet placement produces significant self-weight bending stress, whereas rigt panel shows how a symmetric layout with a high-power central die induces steep CTE-mismatch gradients and elevated thermo-elastic stress. These examples underscore the need to co-optimise temperature and mechanical stress.}

This paper presents \textbf{STAMP-2.5D: Structural and Thermal Aware Methodology for Placement in 2.5D Integration}, the first automated placement methodology that holistically integrates thermal performance, mechanical stress, and interconnect wirelength optimizations. Our proposed framework utilizes finite element analysis (FEA) simulations using ANSYS to concurrently evaluate thermal and mechanical stress profiles, enabling effective exploration of critical design trade-offs and ensuring robust reliability in advanced packaging. We also discuss how temperature gradients affect stress in the package and, finally, provide a fine-grained analysis of package temperature and stress distributions.
\vspace{-0.1cm}
\section{Literature Review}

Effective thermal-mechanical stress management in 2.5D chiplet architectures remains a pivotal challenge, driving extensive research in both component-level analysis and system-level optimization. Early efforts, such as Jung et al. (2011), aimed at reducing computational complexity in reliability analysis by employing linear superposition of stress tensors to assess through-silicon vias (TSVs) stresses, which provided computational efficiency compared to traditional finite element methods (FEM)\cite{Jung2011}. Nevertheless, this approach was limited to isolated TSV structures, neglecting integrated system-level implications and optimizations. Similarly thermal stresses in copper interwafer interconnects bonded with benzocyclobutene (BCB) are analysed offering insights into interwafer via stability yet without considering full-chip or system-wide interactions\cite{Zhang2006}. More recently, thermal-stress management by integrating design-time and run-time methodologies to minimize thermo-mechanical stresses in 3-D ICs have been ivestigated. Their work emphasized the importance of dynamic thermal gradients but lacked an automated optimization strategy explicitly targeting chiplet placement for simultaneous thermal and mechanical optimization~\cite{Yuan17}.

Multi-scale and multi-fidelity modeling frameworks have emerged as promising techniques to bridge component-level and system-level analyses. Studies have integrated detailed thermal and mechanical analyses using finite element simulations with compact modeling methods, enabling performance predictions across multiple scales from individual components to complete systems~\cite{Wang2018,Choy24}. These approaches effectively reconcile localized stress evaluations with global thermal predictions, yet notably lack automated optimization algorithms necessary for practical placement decisions in chiplet-based systems.

Concurrently, automated placement methodologies have made significant strides in optimizing various design objectives, predominantly thermal performance, interconnect wirelength, and power efficiency. Prominent studies include those leveraging mathematical programming, heuristic-driven optimization, and machine-learning-based methods~\cite{Shi2023Floorplet,Hong2023Chiplet,TAP25D,Duan2023RLPlanner,Zhi2024Chiplet}. For instance, TAP-2.5D by Ma et al. (2023) explicitly introduced thermally-aware placement techniques, emphasizing insertion of spacing to reduce temperatures~\cite{TAP25D}. Chiou et al. (2023) and Deng et al. (2024) similarly utilized heuristic and reinforcement learning-based methods for thermal-aware placement but did not explicitly incorporate mechanical stress as an objective or constraint in their methodologies~\cite{Hong2023Chiplet,Zhi2024Chiplet}. Although these studies represent significant advances in thermal and interconnect optimization, structural reliability and the critical implications of mechanical stress, especially arising from thermal gradients and interposer deformation, have been notably neglected.
There is an critical need for automated placement methodology to jointly optimize chiplet configurations for thermal management, mechanical stress minimization, and interconnect wirelength efficiency.

\section{Methodology}

This section describes the methodology for optimizing thermal, mechanical, and interconnect performance in 2.5D chiplet systems. We use a Simulated Annealing (SA) framework to balance thermal, mechanical, and interconnect objectives, with performance evaluated through ANSYS-based Finite Element Analysis (FEA) simulations. Mechanical reliability is assessed using von Mises stress, while wirelength is estimated to ensure efficient interconnect routing, enabling comprehensive optimization of chiplet-based architectures.
\begin{figure}[ht]
    \includegraphics[width=1.05\linewidth]{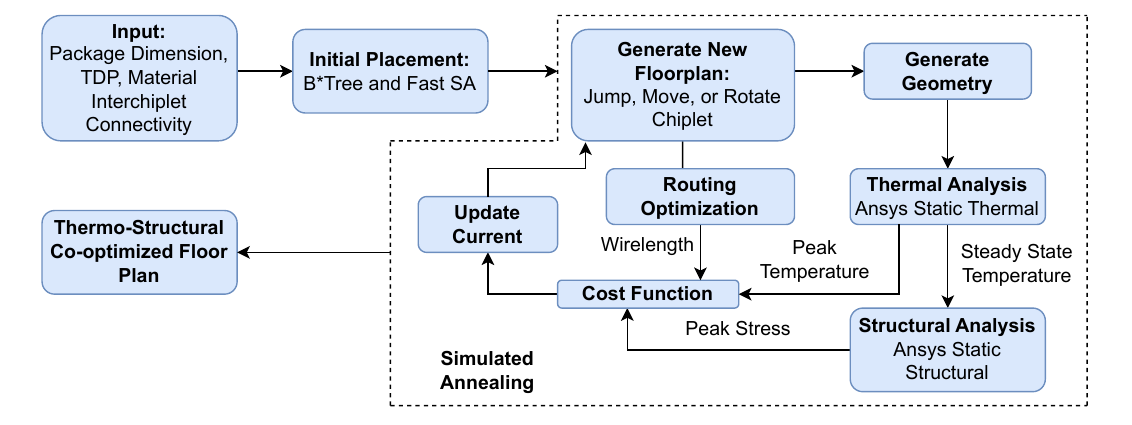}
    \caption{STAMP-2.5D automated placement algorithm overview.}
    \vspace{-0.2cm}
    \label{fig:stamp_algorithm_flow}
    \vspace{-0.3cm}
\end{figure}

\subsection{Simulated Annealing}
Our simulated annealing framework integrates a dynamically adaptive cost function that rigorously balances thermal, mechanical, and dimensional constraints. The process begins by calculating adaptive weights---denoted as \(a\), \(b\), and \(c\)---which adjust based on the current and candidate design parameters, thereby allowing the algorithm to respond sensitively to variations in operating conditions.

Temperature Weight (\(a\)):
Considering 23\(^\circ\)C as the ambient baseline, the temperature weight is set to zero for conditions where both the old and new temperatures are below 75\(^\circ\)C. For higher temperatures, \(a\) is defined as:
\begin{equation}
\begin{split}
a = & \min\Bigl(0.1 + 0.01 \cdot \bigl(\max(T_{\text{old}}, T_{\text{new}}) - 23\bigr),\, 0.5\Bigr) \\
& \times \frac{\max(T_{\text{old}}, T_{\text{new}}) - 60}{40}
\end{split}
\end{equation}

Here, the first term provides a base penalty starting at 0.1, which scales with the temperature’s deviation from the ambient (23\(^\circ\)C), capping at 0.5. The subsequent normalization factor ensures that the weight is appropriately scaled over the interval from 60\(^\circ\)C to 100\(^\circ\)C, capturing the increased significance of temperature beyond nominal operating conditions.

Stress Weight (\(c\)):
The weight for stress is computed to impose a progressively stronger penalty as the system approaches its mechanical limits. This is achieved by incorporating a nonlinear sensitivity term via an exponent:
\begin{equation}
b = \min\left(0.1 + 0.5\left(\frac{\max(\sigma_{\text{old}}, \sigma_{\text{new}})}{\sigma_{\text{max}}}\right)^{1.5},\, 0.5\right)
\end{equation}
represent the stress values of the current and candidate states, while \(\sigma_{\text{max}}\) is the maximum allowable stress. The nonlinearity introduced by the exponent of 1.5 ensures that as the stress approaches its upper limit, even marginal increases are penalized disproportionately, thereby safeguarding the system's structural integrity.

Length Weight (\(b\)):
The length weight is derived as the residual factor needed to balance the overall weight distribution: 
    $c = \left|1 - a - b\right|$

This formulation accounts for the normalized change in length between iterations, ensuring that variations in dimension are factored into the cost function in a balanced manner relative to temperature and stress.
The overall cost for a design is computed as a weighted sum of the normalized deviations in temperature, length, and stress. When normalization ranges are available, the cost functions for the current and candidate solutions are given by: 
\begin{equation}
\begin{aligned}
\text{old\_cost} &= a \cdot T^{\text{old}}_{\text{norm}} + b\cdot \sigma^{\text{old}}_{\text{norm}}  + c \cdot L^{\text{old}}_{\text{norm}} \\
\text{new\_cost} &= a \cdot T^{\text{new}}_{\text{norm}} + b \cdot \sigma^{\text{new}}_{\text{norm}} + c \cdot L^{\text{new}}_{\text{norm}} \\
X_{\text{norm}} &= \frac{X - X_{\text{min}}}{X_{\text{max}} - X_{\text{min}}}
\end{aligned}
\end{equation}
If the normalization ranges are not applicable, a linear formulation is adopted instead. The difference in cost is defined as: 
    $\Delta = -\left(\text{new\_cost} - \text{old\_cost}\right)$.
A positive \(\Delta\) (i.e., a decrease in cost) results in the immediate acceptance of the candidate solution. Conversely, if \(\Delta\) is negative, the candidate solution is accepted with a probability determined by the Boltzmann criterion: 
    $ap =\exp\left(\frac{\Delta}{T}\right)$. 

The $T$ annealing temperature here is different from the package temperature. This probabilistic acceptance allows the algorithm to escape local minima by occasionally accepting solutions with higher costs, particularly during the early stages when \(T\) is high. As the system cools, the likelihood of accepting suboptimal solutions diminishes, guiding the process toward convergence on a near-optimal configuration. The baselines for comparison are Wirelength + Temperature (WT) representing conventional thermal-aware placement and  Wirelength + Stress (WS) optimizing mechanical reliability without thermal coupling. Both use identical SA parameters for fair comparison.
\subsection{Von Mises Stress}
Mechanical reliability in 2.5D chiplet systems is typically assessed using \textit{von Mises stress}, a scalar metric derived from the stress tensor components. Von Mises stress is particularly relevant for evaluating the likelihood of yielding or permanent deformation in materials subjected to complex loading conditions such as integrated circuits\cite{Yuan17,Jung2011}. Mathematically, von Mises stress is defined as:
\begin{equation}
\begin{aligned}
\sigma_{\mathrm{vm}}
  &= \Biggl[
      \underbrace{\tfrac12\!\Bigl((\sigma_x-\sigma_y)^2
      +(\sigma_y-\sigma_z)^2
      +(\sigma_z-\sigma_x)^2\Bigr)}_{\substack{\text{CTE‐mismatch - normal stresses}}}
\\[-1pt]
  &\quad
      +\;\underbrace{3\!\Bigl(\tau_{xy}^2+\tau_{yz}^2+\tau_{zx}^2\Bigr)}_{\substack{\text{self‐weight - shear stresses}}}
     \Biggr]^{1/2}
\end{aligned}
\end{equation}

where $\sigma_{vm}$ represents the von Mises stress, the terms $\sigma_x$, $\sigma_y$, and $\sigma_z$ denote normal stresses in the respective coordinate directions, and $\tau_{xy}$, $\tau_{yz}$, and $\tau_{xz}$ represent the shear stresses. Our approach employs this metric within ANSYS finite element analysis (FEA) to accurately capture structural stress profiles across heterogeneous chiplet arrangements. The FEA model applies both the steady-state temperature field and a gravity load (9.81 ms$^{-2}$) to capture the self-weight induced shear stress by the components towards bending and warpage.

\subsection{Wirelength estimation}
Finding the minimum wirelength for inter-chiplet communication while satisfying bandwidth requirements is an NP-hard problem in multi-chiplet architectures. To accurately estimate interconnect wirelength, we adapted the technique from TAP-2.5D~\cite{TAP25D}, integrating it into our multi-objective optimization framework. Our approach employs a Mixed-Integer Linear Programming (MILP) solver with an objective function that minimizes total wirelength:
$\min \sum_{i,j \in C} d_{ij} \cdot w_{ij}$ 
where $d_{ij}$ represents the distance between chiplets $i$ and $j$, and $w_{ij}$ denotes the number of wires connecting them.\\
The solver takes as input: (1) chiplet placement from our thermo-mechanically aware placement algorithm, (2) microbump resources for inter-chiplet communication, and (3) bandwidth constraints. Flow conservation constraints ensure the net flow from source to sink meets bandwidth requirements, while all intermediate nodes maintain zero net flow. The solver outputs the optimal routing and total wirelength with an average runtime of five seconds, enabling efficient design space exploration.

\subsection{Overview of Automated placement Algorithm}
The proposed STAMP-2.5D methodology leverages a Simulated Annealing (SA) optimization framework to holistically optimize thermal management, structural reliability, and interconnect efficiency in chiplet-based systems. The algorithm initiates from an initial feasible chiplet placement—generated by adapting a widely-used automated placement technique~\cite{Chen2006ModernFB}—and iteratively explores improved configurations through stochastic perturbations.

At each iteration, the algorithm perturbs chiplet positions or orientations to generate candidate solutions, which are evaluated using an integrated cost function accounting for thermal performance, structural stress (von Mises), and interconnect wirelength. Thermal and mechanical stress profiles for candidate configurations are obtained through finite element analysis (FEA) simulations in ANSYS, ensuring accurate stress characterization. Interconnect wirelength is concurrently estimated using a MILP-based solver. Candidate configurations are then probabilistically accepted based on the Simulated Annealing criterion, enabling comprehensive exploration of the solution space and avoidance of local minima.

Fig.~\ref{fig:stamp_algorithm_flow} provides a clear depiction of the iterative STAMP-2.5D placement algorithm and highlights the integrated evaluation process.

\subsection{Modeling}
\begin{figure}
    \includegraphics[width=1.04\linewidth]{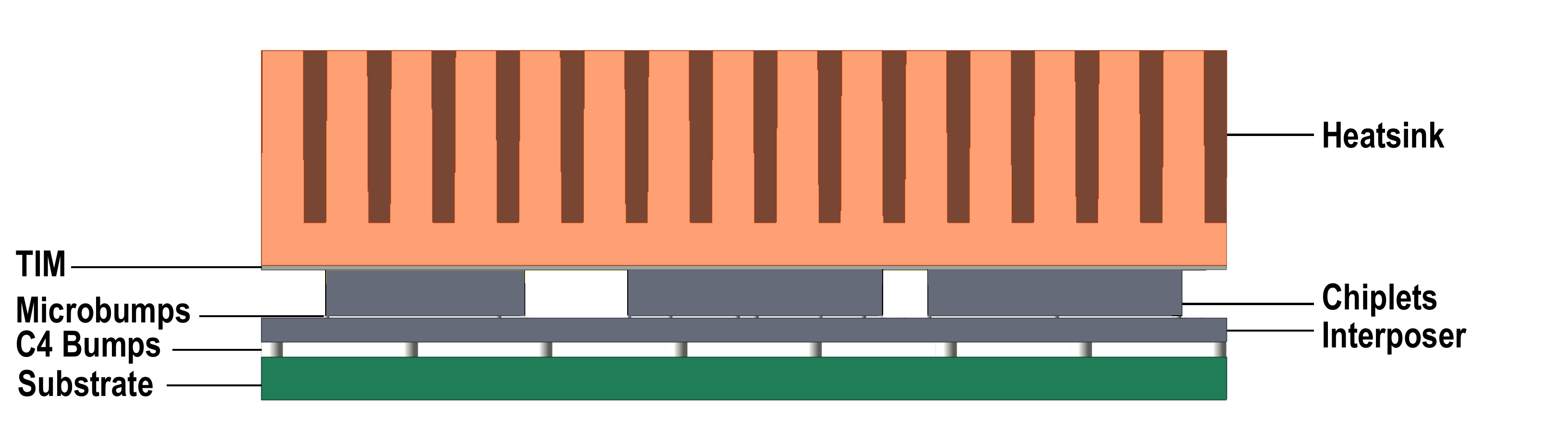}
    \vspace{-0.6cm}
    \caption{Cross-sectional schematic of the 2.5D chiplet integration stack used in STAMP-2.5D modeling. The structure includes key components such as C4 bumps, thermal interface materials (TIM), underfill, interposer, microbumps, chiplets, and a heatsink. This model serves as the foundation for finite element analysis (FEA) simulations, capturing the thermal and mechanical interactions across the layers.}
    \label{fig:modelling}
    \vspace{-0.6cm}
\end{figure}

Our simulation framework employs ANSYS for coupled thermal-structural finite element analysis (FEA) of 2.5D chiplet systems, rigorously capturing multi-physics interactions across a detailed layer stack, as illustrated in Fig.~\ref{fig:modelling}. The stack comprises, from bottom to top, the substrate, C4 bumps, interposer, microbumps, chiplets, thermal interface material (TIM), and the heat sink. Each layer is modeled with its appropriate geometric and physical attributes. The bumps are approximated as cylinders and the heat sink is modeled as a parametric finned structure, with key parameters such as fin height, base thickness and pitch optimized to ensure efficient convective cooling while maintaining mechanical stability. Table~\ref{tab:material_properties} summarizes the materials and their properties for all components.
\begin{table}[ht]
\vspace{-0.3cm}
\centering
\renewcommand{\arraystretch}{1.2}
\setlength{\tabcolsep}{2pt} 
    \caption{Material Properties for Key Components}
\begin{tabular}{cccccc}
\hline                 \textbf{Components} & \textbf{Materials} & \begin{tabular}[c]{@{}l@{}}\textbf{Thermal}\\ \textbf{Conductivity}\\ \textbf{(w/m.k)}\end{tabular} & \begin{tabular}[c]{@{}l@{}}\textbf{CTE}\\ \textbf{(ppm/°C)}\end{tabular} & \begin{tabular}[c]{@{}l@{}}\textbf{Young's }\\ \textbf{Modulus}\\ \textbf{(gpa)}\end{tabular} \\\hline
\textbf{Substrate}  & FR-4                     & 0.3                    & 13.0                   & 20 \\ 
\textbf{C4 Bump}   & Tin-Lead (60-40)          & 50                     & 20.5                   & 20 \\ 
\textbf{$\mu$bump}  & SAC (SnAgCu)              & 50                     & 20                      & 50  \\ 
\textbf{Interposer} & Silicon Anisotropic       & 148                    & 2.6                     & 150\\ 
\textbf{Chiplets}   & Silicon Anisotropic       & 150                   & 3.1                     & 130 \\ 
\textbf{TIM}        & Indium                    & 86                     & 29                      & 10 \\ 
\textbf{Heat Sink}  & Copper Alloy             & 398                   & 16                      & 120\\ 
\textbf{Underfill}  & SiO2                      & 1.4                    & 0.5                     & 80 \\ \hline
\end{tabular}
\vspace{-0.2cm}
\label{tab:material_properties}
\end{table}
A refined mesh is applied in regions exhibiting steep thermal gradients and high stress concentrations—especially near interfaces and power-dense zones—to guarantee numerical precision. 
The simulation is conducted under steady-state conditions with a fixed ambient temperature of 23°C, and forced convective boundary conditions are applied on the heat sink surfaces to replicate realistic active cooling scenarios. By leveraging ANSYS’s multi-physics capabilities, our framework concurrently evaluates heat conduction and the resultant mechanical stresses. Specific experimental parameters, including localized power densities and variations in boundary conditions, are discussed in detail in the subsequent section.

\subsection{Experimental Setup}
\begin{figure*}
    \centering
\includegraphics[width=0.75\linewidth]{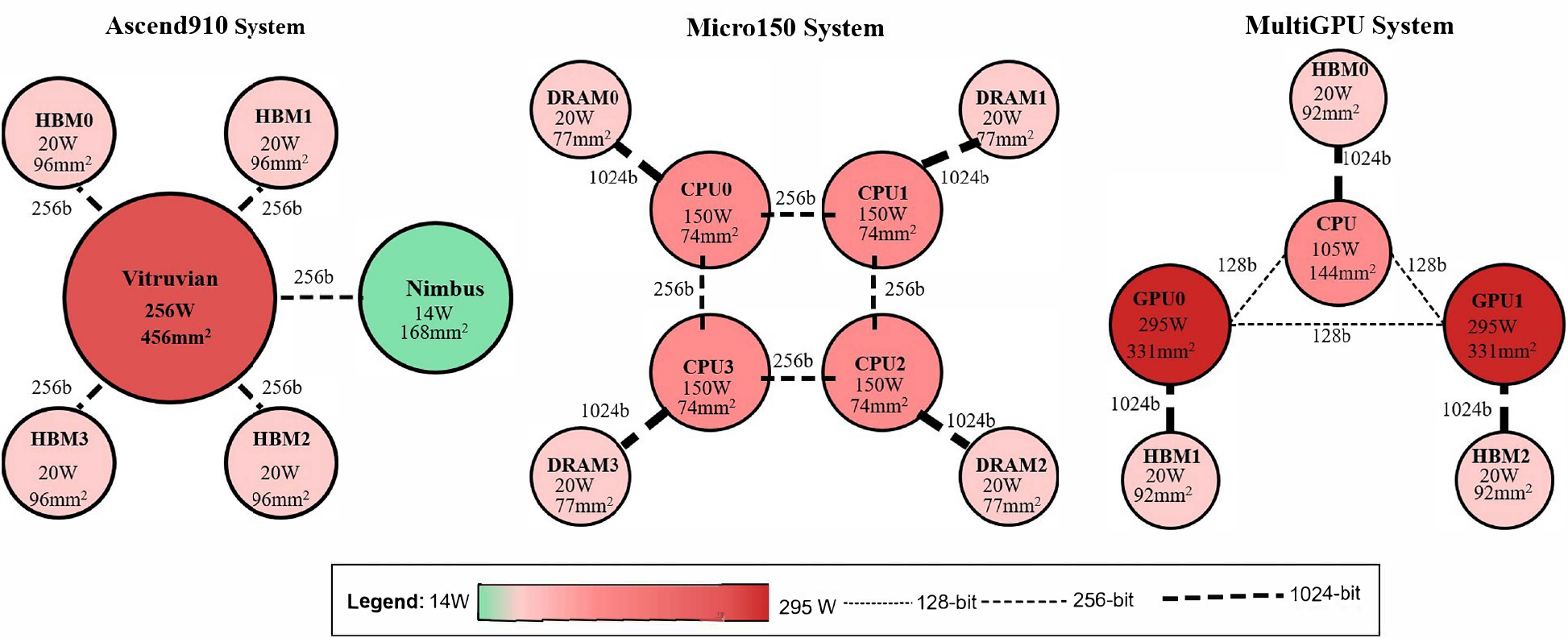}
    \caption{The network topologies, chiplet power and connectivity for Ascend910\cite{HuaweiAscend910}, Micro150\cite{micro150} and MultiGPU\cite{TAP25D} architecture, highlighting the varying power requirements (indicated by color intensity) and bandwidth requirements (indicated by connection line thickness) between components.}
    \label{fig:experimental_setup}
\end{figure*}

Our experimental evaluation utilized three representative architectures: Micro150, Ascend910, and MultiGPU, which cover various chiplet configurations found in research and industry applications, as illustrated in Fig.~\ref{fig:experimental_setup}. We implemented these architectures on a 2.5D package with a silicon interposer of dimensions 45mm × 45mm for ascend910 and 50mm x 50mm for the rest and thickness 100$\mu$m. Each architecture consists of multiple chiplets with varying dimensions and functionalities.
For our thermal-mechanical-aware simulated annealing-based placement experiments, we maintained a consistent chiplet thickness of 150$\mu m$ across all chiplet types. This standardized thickness was chosen to represent typical manufacturing practices for high-performance chiplets. However, in our exploratory studies discussed in the results section, we varied the chiplet thickness to investigate the impact on thermal and mechanical performance. For thermal simulation, we applied non-uniform power density distributions across chiplets detailed in Fig.~\ref{fig:experimental_setup}.
The boundary conditions for our thermal simulations included an ambient temperature of 23°C with forced convective cooling applied to the heat sink. We used architecture-specific heat transfer coefficients to model different cooling scenarios: 580 W/m²·K for Ascend910, 720 W/m²·K for Micro150, and 950 W/m²·K for MultiGPU \cite{submerConvection}. These values reflect the different thermal requirements of each architecture, with higher-power systems requiring more effective cooling solutions. The bottom surface of the package substrate was set to a natural convection boundary condition with a heat transfer coefficient of 10 W/m²·K.

All finite element analyses were run in ANSYS Mechanical for static thermomechanical conditions. We used a medium mesh of roughly forty thousand elements with local refinement near interfaces and strong thermal gradients, and we chose the coarsest mesh whose temperature predictions were negligible by additional refinement.

\textcolor{black}{Our optimization algorithm parameters included an initial annealing temperature of 1.0, a cooling rate of 0.9, and 45 iterations per temperature level for Ascend910 and 50 for Micro150 and MultiGPU, with convergence at an annealing temperature of $\leq$ 0.01. This resulted in approximately 60.5 hours for Ascend910 and 67 hours for Micro150 and MultiGPU. With 5 independent runs for statistical validation, complete optimization requires 300-335 hours per architecture.}
\vspace{-0.6cm}
\section{Result and Analysis}
We present experimental results from our thermal-mechanical aware placement algorithm and analyze the critical relationship between temperature gradients and von Mises stresses in heterogeneous chiplet systems. We compare our results against thermal-wirelength (TW) and thermal-stress (TS) co-optimisation only. The Thermal-Wirelength only optimisation is similar to \cite{TAP25D} except that the geometry and consequently temperature prediction is performed using Ansys and not HotSpot. 
\subsection{Thermal mechanical aware automated placement}
\begin{figure}[t]
    \centering 
    \includegraphics[width=\linewidth]{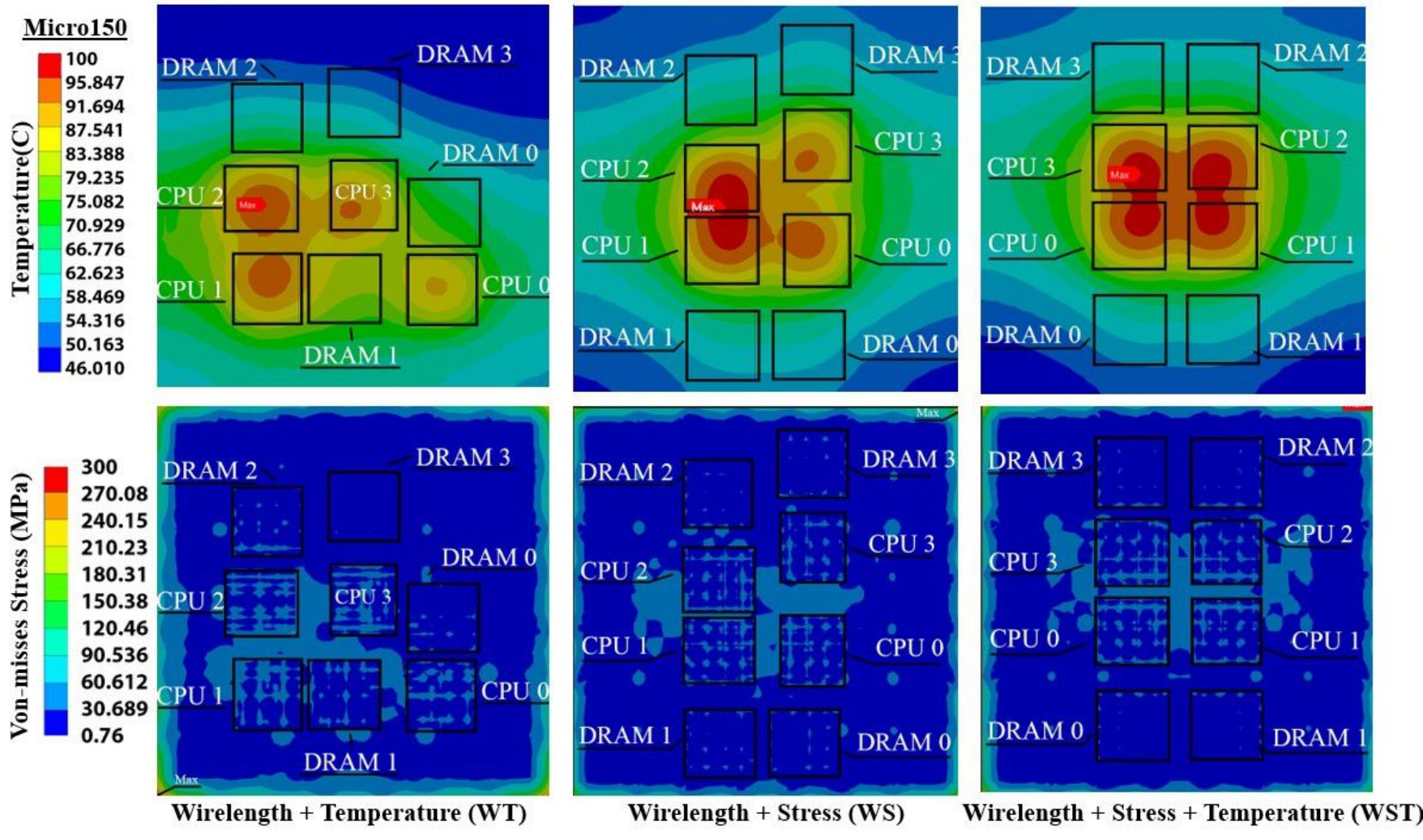}
    \vspace{-0.7cm}
    \caption{Temperature and von Mises stress profile of the placement generated by the algorithm considering different cost functions (x-axis) for \textbf{Micro150 architecture}}
    \label{fig:micro150}
    \vspace{-0.50cm}
\end{figure}

 We quantify temperature distributions, von Mises stress, and total interconnect wirelength through Ansys-based finite element analysis (FEA) simulations. Table~\ref{tab:architecture_metrics} summarizes our results, clearly demonstrating the advantages of STAMP-2.5D compared to conventional thermal placement methodologies, while highlighting essential trade-offs among thermal management, structural reliability and connectivity efficiency.\\
Numerical results reveal compelling trade-offs. In the Micro150 architecture, our integrated approach notably reduces von Mises stress by 19.8\%. This significant stress reduction is primarily driven by a symmetric chiplet placement strategy, which ensures uniform weight distribution across the interposer. Despite a modest temperature increase of only 2.97\%, our method simultaneously achieves a wirelength reduction of 19.3\%. The stress maps presented particularly in Fig.~\ref{fig:micro150} clearly illustrate how symmetric placements lead to more balanced mechanical loads, contrasting sharply with the concentrated stress points typical of temperature-only optimization. \textcolor{black}{Because self-weight loading is included in the structural model, symmetric placement lowers bending moments on the interposer, thereby reducing peak $\sigma_{vm}$ even when peak $T$ rises slightly. This balanced load distribution is critical in preventing localized joint failures and reducing interposer cracking during thermal cycling.}

\begin{figure}[t]
    \centering 
    \includegraphics[width=1\linewidth]{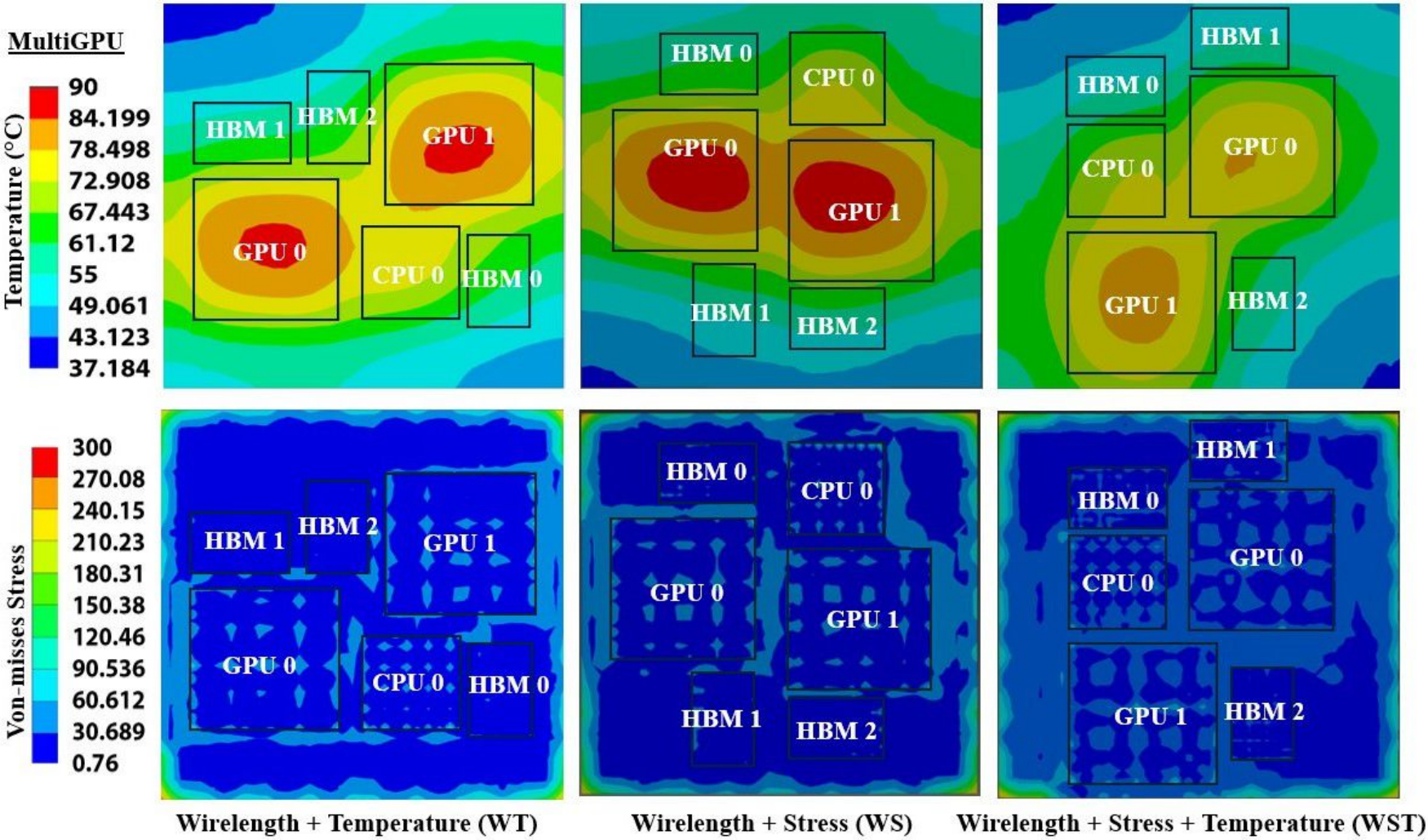}
    \vspace{-0.7cm}
    \caption{Temperature and von Mises stress profile of the placement generated by the algorithm considering different cost functions (x-axis) for \textbf{MultiGPU architecture}}
    \label{fig:multigpu}
    \vspace{-0.4cm}
\end{figure}

In parallel, the MultiGPU architecture demonstrates similar benefits. Here, our method achieves a 7.8\% stress reduction, accompanied by a negligible temperature increase of just 0.39\%, while improving the wirelength by 5.2\%. The balanced placement proves particularly advantageous for managing the high power densities typical of GPU components within multi-GPU systems. Stress maps provided in Fig.~\ref{fig:multigpu} illustrate a clear reduction in localized stress near power-intensive GPU components. By more uniformly distributing thermal gradients across the interposer, our combined approach significantly reduces localized stress concentrations. This controlled management of thermal gradients effectively mitigates potential mechanical vulnerabilities caused by uneven expansion and contraction, as analyzed further in subsequent sections.

\begin{figure}[t]
    \centering 
    \includegraphics[width=1\linewidth]{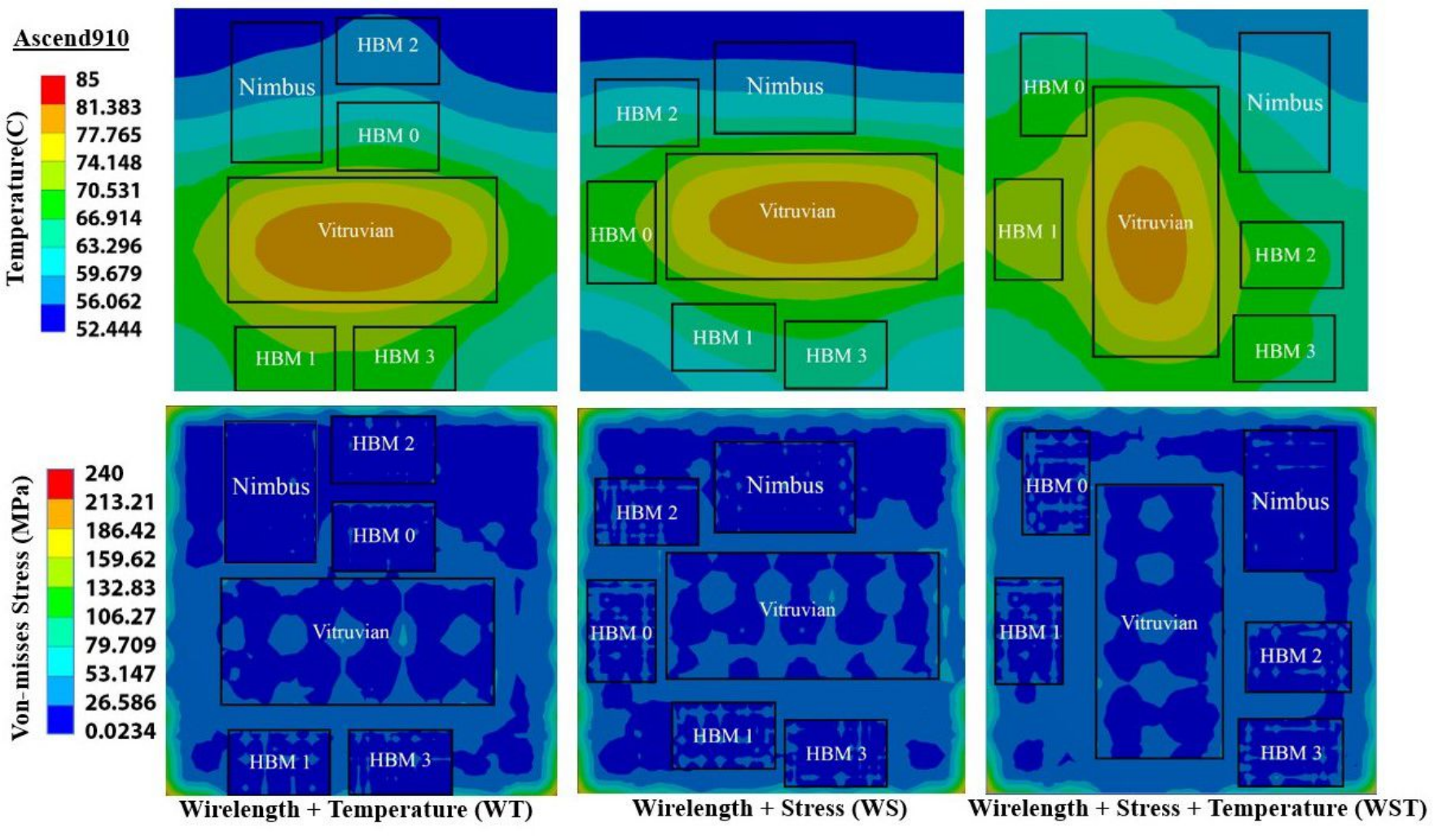}
    \vspace{-0.7cm}
    \caption{Temperature and von Mises stress profile of the placement generated by the algorithm considering different cost functions (x-axis) for \textbf{Ascend910 architecture}}
    \label{fig:ascend910}
    \vspace{-0.5 cm}
\end{figure}

Consistent improvements are observed in the Ascend910 architecture, where our optimization strategy delivers dual advantages: a 4.5\% reduction in von Mises stress alongside a 2.5\% decrease in peak temperature. The temperature contour maps in Fig.~\ref{fig:ascend910} vividly demonstrate how structurally-aware placement effectively redistributes thermal concentrations around the critical vitruvian component. Specifically, the sharp orange-red hotspot prominent in temperature-only optimization is transformed into smoother, more gradual temperature transitions with the combined method. This deliberate spreading of thermal gradients significantly enhances mechanical stability by minimizing localized expansion differentials and concurrently boosts overall thermal efficiency by utilizing a greater portion of the interposer’s surface area for effective heat dissipation. The STAMP-2.5D methodology thus represents not merely an optimal middle ground, but in most cases a superior solution: it avoids the high mechanical stress of temperature-only optimization, circumvents the elevated temperatures of stress-only approaches, and provides more reasonable wirelength than single-metric optimizations—ultimately delivering a balanced solution that simultaneously addresses thermal management, structural reliability, and interconnect efficiency in chiplet integration.
\begin{table}[ht]
    \centering
    \vspace{-0.3 cm}\caption{Performance metrics for different architectures under various cost functions. WT: Wirelength + Temperature, WS: Wirelength + Stress, WST: Wirelength, Stress + Temperature.}\renewcommand{\arraystretch}{1.2}
    \setlength{\tabcolsep}{2pt} 
    \begin{tabular}{ccccc}
        \hline
        \multirow{2}{*}{\textbf{Architecture}} & \multirow{2}{*}{\textbf{Cost Function}} & \textbf{Temperature} & \textbf{Stress} & \textbf{Wirelength} \\
         & & (\textdegree C) & (MPa) & (mm) \\

        \hline
        \multirow{3}{*}{\textbf{Ascend910}} 
        & WT & 81.06 & 232.63 & 27200  \\
        & WS & 81.21 & 226.89 & 25895 \\
        & WST & 79.04 & 222.14 & 27787 \\
        \hline
        \multirow{3}{*}{\textbf{MultiGPU}} 
        & WT & 85.69 & 280.82 & 82680 \\
        & WS & 87.07 & 275.85 & 98794 \\
        & WST & 86.04 & 258.97 & 78358 \\
        \hline
        \multirow{3}{*}{\textbf{Micro150}} 
        & WT & 95.24 & 291.02 & 104314  \\
        & WS & 99.04 & 224.00 & 93251 \\
        & WST & 98.07 & 233.47 & 84227 \\
        \hline
    \end{tabular}
    \vspace{-0.6cm}
\label{tab:architecture_metrics}
\end{table}

\textcolor{black}{\subsection{Thermal Gradients and Structural Integrity Analysis}} 
\begin{table*}[h]
    \centering
    \caption{Thermal Gradient Analysis for Different Architectures. WT: Wirelength + Temperature, WS: Wirelength + Stress, WST: Wirelength, Stress + Temperature.}
    \label{tab:metrics}
    \vspace{-0.2cm}\renewcommand{\arraystretch}{1.2}
    \setlength{\tabcolsep}{4pt} 
    \resizebox{\textwidth}{!}{
    \begin{tabular}{l|c|c|c|c|c|c|c|c|c}
        \hline
        \multirow{2}{*}{\textbf{Metrics}} & \multicolumn{3}{c|}{\textbf{Ascend910}} & \multicolumn{3}{c|}{\textbf{Micro150}} & \multicolumn{3}{c}{\textbf{Multi-GPU}} \\
        \cline{2-10} 
        & WT & WS & WST
        & WT & WS & WST 
        & WT & WS & WST \\
        \hline
        Mean Gradient $({}^{\circ}\mathrm{C}\,\mathrm{mm}^{-1})$ & 871.56 & 968.90 & 774.59 & 1521.42 & 1546.38 & 1603.56 & 1457.60 & 1479.46 & 1322.73 \\
        
        Thermal Gradient Standard Deviation $({}^{\circ}\mathrm{C}\,\mathrm{mm}^{-1})$ & 487.60 & 503.18 & 444.88 & 956.72 & 1096.99 & 1202.41 & 879.24 & 819.58 & 735.29 \\        
        
        Max Gradient $({}^{\circ}\mathrm{C}\,\mathrm{mm}^{-1})$ & 9286.08 & 6557.75 & 7384.52 & 16866.98 & 19009.43 & 18333.74 & 17955.89 & 15318.53 & 11880.72 \\
        
        Temperature-Stress Correlation Coefficient (T-S) & -0.115 & -0.171 & -0.215 & -0.018 & -0.089 & -0.042 & -0.158 & -0.173 & -0.176 \\
        
        Gradient-Stress Correlation Coefficient (G-S) & -0.252 & -0.236 & -0.246 & -0.256 & -0.173 & -0.173 & -0.217 & -0.220 & -0.222 \\
        \hline
    \end{tabular}
    }
    \vspace{-0.5cm}
\end{table*}
The fundamental relationship $\sigma \approx E \alpha \Delta T$ shows thermal expansion induces stress proportional to temperature rise. At material interfaces, this becomes $\sigma \propto (\alpha_i - \alpha_j) \Delta T$, where CTE mismatches amplify stress. However, stress generation in 2.5D chiplet systems involves multiple loading mechanisms beyond thermal expansion alone.

Our analysis confirms that temperature gradients, rather than absolute temperatures, are the primary driver of thermally-induced mechanical stress. As shown in Table~\ref{tab:metrics}, the gradient-stress correlation coefficients (G-S Corr: $-0.173$ to $-0.285$) are consistently stronger than temperature-stress correlations (T-S Corr: $-0.018$ to $-0.215$) across all architectures and optimization approaches. This validates that spatial temperature variations create the thermal expansion mismatches that generate mechanical stress.

\begin{figure}
    \centering
    \includegraphics[width=\linewidth]{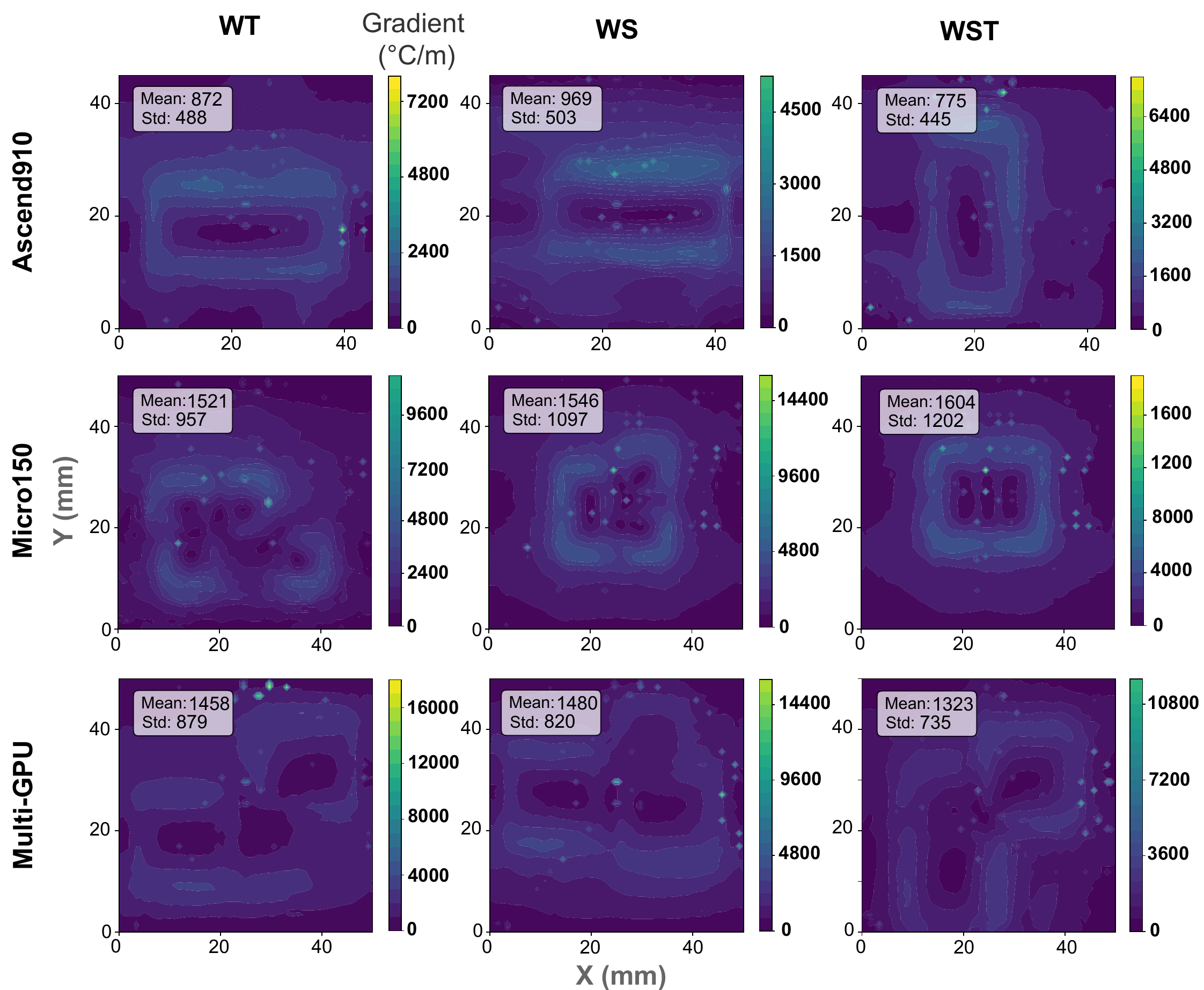}
    \vspace{-0.6cm}
    \caption{Temperature Gradient for MultiGPU, Micro150 and Ascend architecure}
    \label{fig:temp_grad}
    \vspace{-0.6cm}
\end{figure}

Including temperature and stress in the objective leads the SA framework to reduce thermal gradients and mechanical loading asymmetries. Consequently, placements tend to show reduced co-location of temperature and stress hotspots, although full decoupling is not guaranteed. The stress-aware optimization strategy consistently produces more uniform temperature gradient distributions across architectures, as illustrated in Fig.~\ref{fig:temp_grad}, which shows temperature gradient heat maps for all optimization approaches. In Ascend910, our integrated WST approach significantly reduces gradient standard deviation ($444.88$ vs $501.60$, $11\%$ reduction) and maximum gradient ($7384.52$ vs $9286.08$, $20\%$ reduction) as detailed in Table~\ref{tab:metrics}. MultiGPU demonstrates similar improvements with gradient standard deviation reducing from $808.24$ to $738.29$ ($9\%$ reduction) and maximum gradient decreasing from $17955.89$ to $11880.72$ ($34\%$ reduction). These gradient reductions, clearly visible in the gradient maps of Fig.~\ref{fig:temp_grad}, directly correlate with the observed stress improvements while maintaining effective thermal management.

The Micro150 architecture provides particularly compelling evidence for multi-physics stress generation beyond thermal effects. Despite thermal-only optimization (TW) achieving relatively controlled thermal gradients (Std Grad: $1042.72$, Table~\ref{tab:metrics}), it produces high absolute stress values due to asymmetric chiplet placement that creates unbalanced self-weight loading across the interposer. This asymmetric mass distribution generates bending moments independent of thermal effects, demonstrating why comprehensive thermal-mechanical co-optimization is essential for architectures with non-uniform chiplet arrangements. 
\subsection{Impact of Intra-Chiplet Power Density on Thermal-Mechanical Reliability}
\begin{figure}
    \centering
    \includegraphics[width=\linewidth]{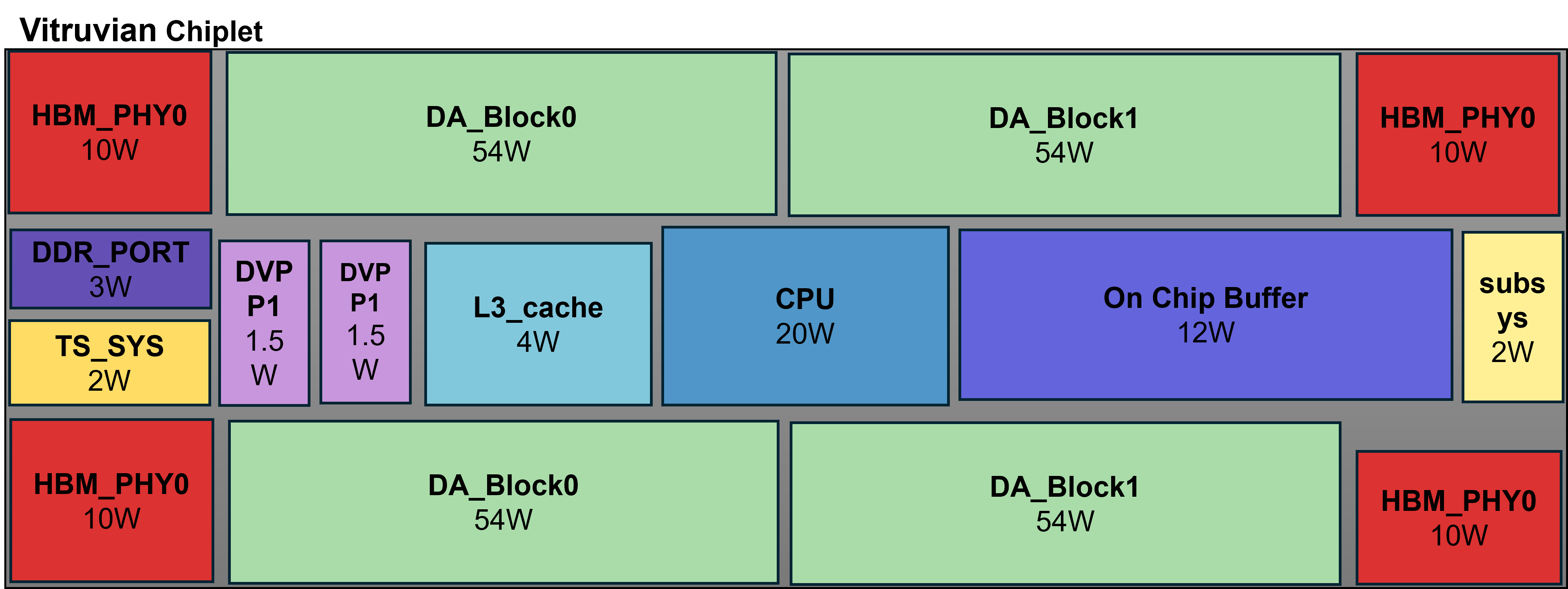}
    \vspace{-0.7 cm}
    \caption{Estimated placement for vitruvian chiplet in Ascend910 architecture\protect \footnotemark{}} 
    \label{fig:intra_chip}
    \vspace{-0.5cm}
\end{figure}

To assess the effect of finer-grained power variation on placement quality, we take the final STAMP-2.5D placements generated under each objective set—wirelength+temperature (WT), wirelength+stress (WS) and the full multi-objective (WST) for the Vitruvian chiplet in the Ascend910 system and recompute their steady-state temperature and von Mises stress fields for a spatially varying heat-flux profile derived from the block-level power map of the Vitruvian compute die (Figure~\ref{fig:intra_chip})\cite{Davinci}. By holding placement constant and only changing the power input, this study isolates how intra-die hotspots influence package‐level thermal and mechanical behavior without retraining or retuning the optimiser.

\begin{figure}[h]
\vspace{-0.5cm}
    \centering    \includegraphics[width=0.8\linewidth]{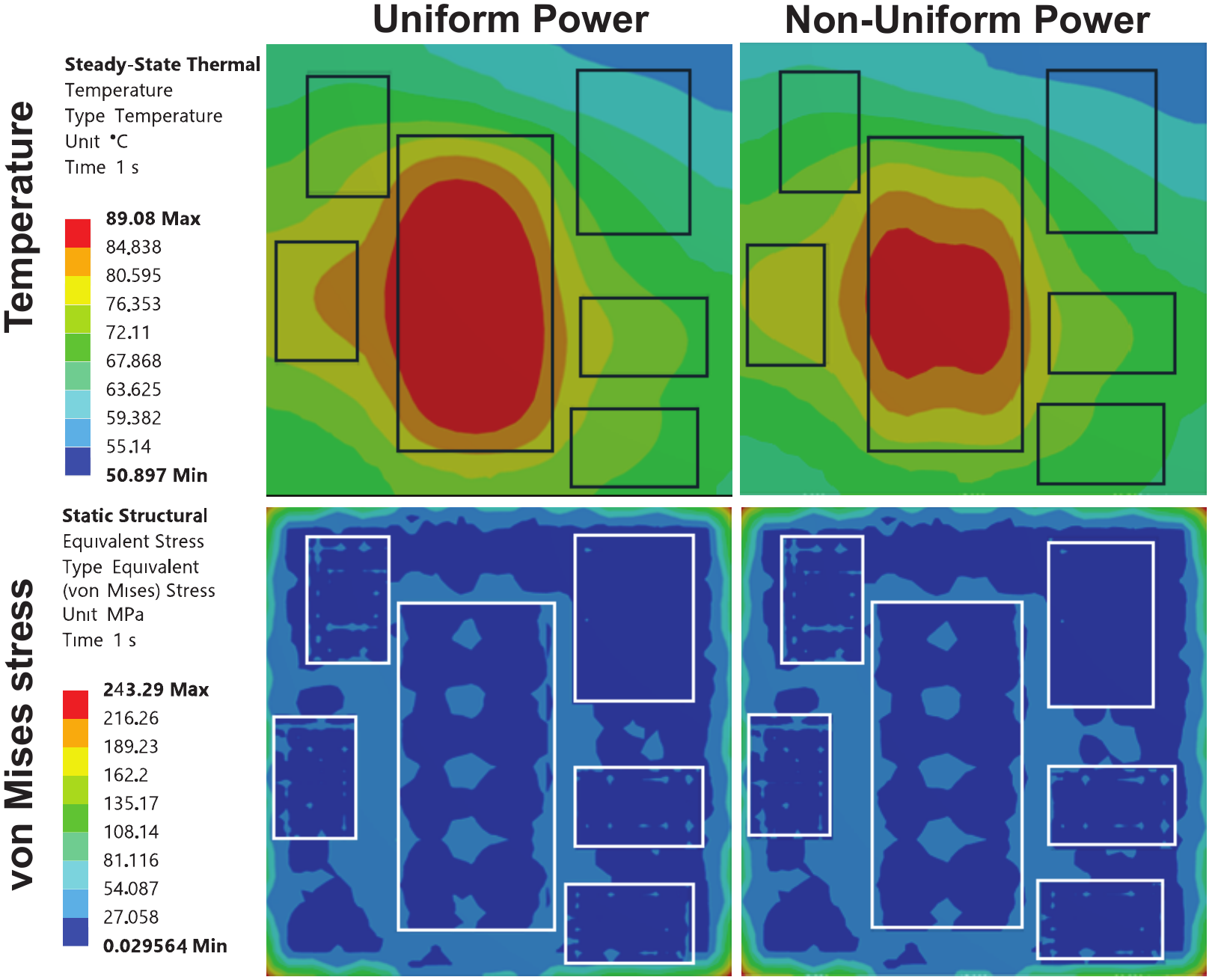}
    \vspace{-0.3cm}
    \caption{Thermal and Stress map for coarse (left) and fine grained (right) power maps for vitruvian chiplet for Ascend910}
    \label{fig:intra_result}
    \vspace{-0.5cm}
\end{figure}

\begin{table}[ht]
\centering
\caption{Peak temperature and peak von Mises stress for different optimization configuration (WT, WS, WST) under uniform and non‐uniform intra‐chiplet power distributions.}
\begin{tabular}{l  cc  cc}
\toprule
Configuration & \multicolumn{2}{c}{Uniform} & \multicolumn{2}{c}{Non-uniform} \\
\cmidrule(lr){2-3} \cmidrule(lr){4-5}
 & Temp (°C) & Stress (MPa) & Temp (°C) & Stress (MPa) \\
\midrule
WT  &  90.99 & 256.98 &  91.17 & 258.82 \\
WS  &  91.27 & 249.91 &  91.48 & 251.45 \\
WST &  87.63 & 242.65 &  89.08 & 243.29 \\
\bottomrule
\end{tabular}
\vspace{-0.2cm}
\label{tab:peak_by_config}
\end{table}
Introducing the non-uniform hotspot map raises absolute peak temperatures and stresses only marginally (Table~\ref{tab:peak_by_config}). Crucially, the relative ranking of the three placement modes remains identical under both power assumptions: WST produces the lowest peak temperature and $\sigma_{vm}$, followed by WT, then WS. Figure~\ref{fig:intra_result} shows the temperature and stress contours for the WST placement under the hotspot model; WT and WS follow the similar trends. Although these findings confirm that STAMP-2.5D’s optimal placements are robust to realistic intra-chiplet power variation, implying that a simple average-power model suffices for early‐stage thermal–mechanical placement, a more granular examination of their transient behaviors remains as planned future work to further validate and refine the methodology.
 \setcounter{footnote}{-1}

\footnotetext{the vitruvian’s thermal design power (TDP) for the placement is estimated at 300Ws—different from the 256W TDP assumed during placement—our motivation is to examine relative variations, not absolute temperature or stress magnitudes}
\section{Conclusion and future work }
In this paper, we introduced an automated placement methodology that reduces thermally-induced structural stress in chiplet packages while optimizing wirelength with minimal temperature impact. Our results emphasize the importance of uniform temperature gradients for structural reliability. The methodology presented here establishes a foundational framework suitable for extensive future explorations, both component-specific and in the broader context of structural reliability within heterogeneous chiplet architectures. In the current STAMP-2.5D setup, high-fidelity ANSYS evaluation at every iteration represents the primary computational bottleneck. As future work, we aim to develop a multi-fidelity optimization framework to reduce runtime without compromising accuracy. In conclusion, we provide a foundational methodology for design space exploration. Additional constraints (signal integrity, power integrity, cost) can be seamlessly incorporated to build a unified, multi-objective optimization framework.
\vspace{-0.4cm}
\bibliographystyle{ieeetr}
\bibliography{ref.bib}
\end{document}